\documentclass[12pt]{article}
\usepackage{venice}
\usepackage{epsfig}

\begin{document}

\renewcommand{\thefootnote}{\alph{footnote}}
  
\title{NEUTRINO MASS AND NEUTRINOLESS DOUBLE BETA DECAY}

\author{ PETR VOGEL}

\address{Kellogg Rad. Lab. 106-38, Caltech,\\
 Pasadena, CA 91125, USA, and\\
Max-Planck-Institut fuer Kernphysik, \\
D-69029 Heidelberg, Germany \\
 {\rm E-mail: pxv@caltech.edu} }
 
 \vspace{1cm}
 
Talk at the ``Neutrino Oscillations in Venice" workshop, Venice, April 2008

\abstract{The motivation for the search for $0\nu\beta\beta$ decay is briefly reviewed. It is stressed
that the exchange of light Majorana neutrinos is not the only possible mechanism of the decay.
The link between lepton number and lepton flavor violation is described and its
role in elucidating the $0\nu\beta\beta$-decay mechanism is discussed. The main topic of the
talk is the evaluation of the nuclear matrix elements and their uncertainty. Various physics effects
that influence the value of the matrix elements are described and the results of the two main methods,
the quasiparticle random phase approximation and the nuclear shell model, are compared. }
   
\normalsize\baselineskip=15pt

\section{Introduction}
In the last decade neutrino
oscillation experiments have convincingly and triumphantly
shown, using both the natural and manmade neutrino sources,
that neutrinos have a finite mass and that the lepton flavor is not
a conserved quantity. These results opened the door to what is often called
the ``Physics Beyond the
Standard Model". In other words, accommodating these findings into a consistent
scenario requires generalization of the Standard Model of electroweak 
interactions that postulates that neutrinos are massless and that consequently
lepton flavor, and naturally, also the total lepton number, are  conserved quantities.

\begin{figure}[htb]
%\centerline{\psfig{file=rv-fig1.eps,width=3.8cm}}
\centerline{\psfig{file=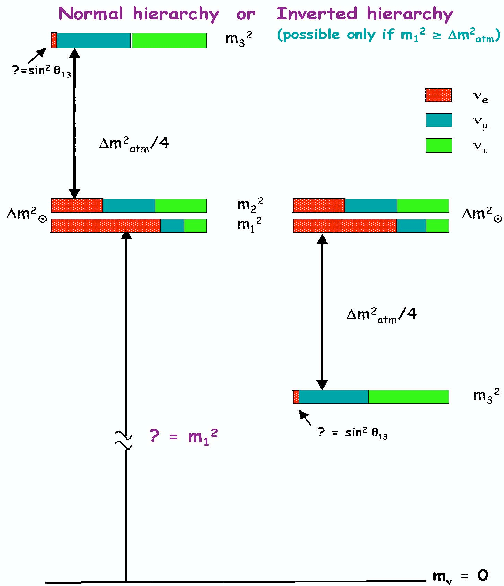,width=9.0cm}}
% \mbox{\epsfig{figure=osc.pdf,width=9.0cm}}
\caption{ Schematic illustration  of the decomposition
of the neutrino mass eigenstates $\nu_i$ in terms of the flavor eigenstates.
The two hierarchies cannot be, at this time, distinguished. The small admixture
of $\nu_e$ into $\nu_3$ is an upper limit, and the mass square of the neutrino 
$\nu_1$, the quantity $m_1^2$, remains unknown. }
\label{fig_osc}
\end{figure}

The present results of the oscillation experiments are summarized in Fig. \ref{fig_osc}
that shows the decomposition of the flavor eigenstates neutrinos $\nu_e, \nu_{\mu}$ and
$\nu_{\tau}$ into the mass eigenstates $\nu_1, \nu_2$ and $\nu_3$.   
An upper limit on the masses of all
active neutrinos $\sim$ 2 - 3 eV can be derived from the combination of
analysis of the tritium beta-decay experiments and the neutrino
oscillation experiments. Combining these constraints, masses of at least two
(out of the total of three active) neutrinos are bracketted by 10 meV $\le m_{\nu} \le$
2 - 3 eV.

Therefore, neutrino masses are  six or more orders of
magnitude smaller than the masses of the other fermions. Moreover,
the pattern of masses, i.e. the mass ratios of neutrinos, is rather different
(even though it remains largely unknown) than the pattern of masses
of the up- or down-type quarks or charged leptons. All of these facts
suggest that, perhaps, the origin of the neutrino mass is different
than the origin (which is still not well understood) of the masses of the other 
charged fermions.    

The smallness of the neutrino masses can be understood following the
finding of Weinberg \cite{Wei79} who pointed out almost thirty year ago 
that there exists only one
lowest order (dimension 5, suppressed by only one inverse power of the
corresponding high energy scale $\Lambda$) gauge-invariant operator given the content
of the standard model. After the spontaneous symmetry
breaking when the Higgs acquires vacuum expectation value that operator
represents the neutrino Majorana mass which violates the total lepton number
conservation law by two units,
\begin{equation}
{\cal L}^{(M)} = \frac{C^{(5)}}{\Lambda} \frac{ v^2}{2} (\bar{\nu}^c \nu) + h.c. ~~,
\end{equation}
where $v \sim$ 250 GeV and $C^{(5)}$ is expected to be of the order of unity.
For sufficiently large scale $\Lambda$ neutrinos masses are arbitrarily small.

The most popular explanation of the smallness of neutrino mass is the see-saw mechanism,
 which is also roughly thirty years old \cite{seesaw}. In it, the existence of heavy right-handed
 neutrinos $N_R$ is postulated, and by diagonalizing the corresponding mass matrix
 one arrives at the formula
 \begin{equation}
 m_{\nu} = \frac{m_D^2}{M_N}
 \end{equation}
 where the Dirac mass $m_D$ is expected to be a typical charged fermion mass and $M_N$ is
 the Majorana mass of the heavy neutrinos $N_R$. Again, the small mass of the standard neutrino
 is related to the large mass of the heavy right-handed partner. Requiring that $m_{\nu}$ is of the
 order of 0.1 eV means that $M_N$ (or $\Lambda$) is $\sim 10^{14-15}$ GeV, i.e. near the GUT
 scale. That makes this template scenario particularly attractive.
 
 Clearly, one cannot reach such high energy scale experimentally. But these scenarios imply that 
 neutrinos are Majorana particles, and consequently that the total lepton number should not
 be conserved. Hence the tests of the lepton number conservation acquires a fundamental
 importance.
  
There are various ways to test whether the total lepton number is conserved or not.
Examples of the potentially lepton number violating (LNV)
processes with important limits are
\begin{eqnarray}
&& (Z,A) \rightarrow (Z+2,A) + 2e^-; {~\rm halflife~ > ~10^{25} ~ years} 
\nonumber \\
&& \mu^- + (Z,A)  \rightarrow  e^+  + (Z-2,A); {~\rm exp.~ branching~ ratio} \le 10^{-12} ~,
\nonumber \\
&&  K^+  \rightarrow   \mu^+ \mu^+ \pi^-;  {~\rm exp. ~branching ~ratio} \le 3 \times 10^{-9} ~,
  \nonumber \\
&&  \bar{\nu}_e {\rm ~emission~ from~ the~ Sun};  {~\rm exp.~ branching~ ratio} \le 10^{-4} ~.
\end{eqnarray}  
However, detailed analysis suggests that the study of the $0\nu\beta\beta$ decay,
the first on the list above, is by far the
most sensitive test of LNV. In simple terms this is caused by the amount of tries one can make.
A 100 kg $0\nu\beta\beta$ decay source contains $\sim 10^{27}$ nuclei
that can be observed for a long time (several years). This can be contrasted
with the possibilities of first producing muons or kaons, and then searching for the unusual
decay channels. The Fermilab accelerators, for example, produce $\sim 10^{20}$ protons on target
per year in their beams and thus correspondingly smaller numbers of muons or kaons.  

\section{Basic considerations}
Double beta decay ($\beta\beta$) is a nuclear transition 
$(Z,A) \rightarrow (Z+2,A)$ in which two neutrons
bound in a nucleus  are  simultaneously transformed into two protons plus two
electrons (and possibly other light neutral particles). This transition is
possible and potentially observable because 
nuclei with even $Z$ and $N$ are more bound than the odd-odd nuclei with
the same $A = N + Z$. 
Analogous transition of two protons into two neutrons
are also, in principle, possible
in several nuclei, but phase space considerations give preference to the former.

There are two basic modes of the $\beta\beta$ decay. In the two-neutrino mode ($2\nu\beta\beta$)
there are 2 $\bar{\nu}_e$ emitted together with the 2 $e^-$. 
It is just an ordinary beta decay of two bound neutrons occurring simultaneously, since
the sequential decays are forbidden by the energy conservation law. 
For this mode, clearly,
the lepton number is conserved and
this mode of decay is allowed in the standard model of electroweak interactions.
It has been repeatedly observed in a number of cases and proceeds with a typical half-life
of $\sim 10^{19-20}$years.
In contrast, in the neutrinoless
mode ($0\nu\beta\beta$) only the 2$e^-$ are emitted and nothing else. 
That mode clearly violates the law
of lepton number conservation and is forbidded in the standard model. Hence, its observation
would be a signal of a "new physics". 

One can  separate the two modes experimentally
by measuring the sum energy of the emitted electrons with a good energy
resolution, even if the decay rate for the $0\nu\beta\beta$ mode is much smaller
than for the   $2\nu\beta\beta$ mode. This is possible since in the $2\nu\beta\beta$ mode
the spectrum is continuous, peaked below the $Q/2$ value of the sum kinetic energy, while
in the   $0\nu\beta\beta$ mode the spectrum is a $\delta-$function at $E_1 + E_2 = Q$,
smeared only by the experimental energy resolution (the nuclear recoil is always
negligibly small). 

\begin{figure}
%\centerline{\psfig{file=rv-fig1.eps,width=3.8cm}}
\centerline{\psfig{file=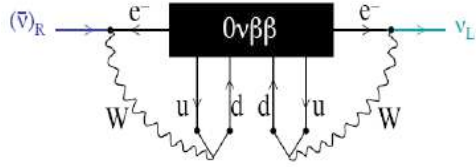,width=7.0cm}}
\caption{ By adding loops involving only standard weak interaction processes the
$0\nu\beta\beta$ decay amplitude (the black box)
implies the existence of the Majorana neutrino mass.  }
\label{fig_SV}
\end{figure}

The existence of the $0\nu\beta\beta$ decay
would mean that on the elementary particle level a six fermion 
lepton number violating amplitude
transforming two $d$ quarks into two $u$ quarks and two electrons
is nonvanishing.  As was first pointed out by Schechter and Valle\cite{SV82} 
more than twenty five years ago,
this fact alone would guarantee that neutrinos are massive 
Majorana fermions (see Fig. \ref{fig_SV}). This qualitative statement (or theorem),
however, does not in general allow us to deduce the magnitude of the neutrino mass
once the rate of the $0\nu\beta\beta$ decay have been determined.
It is important to stress, however, that quite generally an observation of {\bf any} total
lepton number violating process, not only of the $0\nu\beta\beta$ decay, would necessarily 
imply that neutrinos are massive Majorana fermions.

\section{Mechanism of the $0\nu\beta\beta$ decay}

The rather conservative
assumption of how the $0\nu\beta\beta$ decay proceeds is 
to believe that the only possible way
the $0\nu\beta\beta$ decay can occur is through the exchange
of a virtual light, but massive, 
Majorana neutrino 
(the neutrinos $\nu_1$, $\nu_2$ and $\nu_3$ of the first section)
between the two nucleons undergoing the transition,
and that these neutrinos interact by the standard left-handed weak currents. 

If we accept this we can relate the   $0\nu\beta\beta$-decay  rate
to a quantity containing information about  the absolute neutrino mass. 
With these caveats that relation 
can be expressed by a well known formula
\begin{equation}
\frac{1}{T_{1/2}^{0\nu}} = G^{0\nu}(Q,Z) |M^{0\nu}|^2 \langle m_{\beta\beta} \rangle^2 ~,
\label{eq_rate}
\end{equation}
where $G^{0\nu}(Q,Z)$ is a phase space factor that depends on the transition $Q$ value and through
the Coulomb effect on the emitted electrons on the nuclear charge, and that can be
easily and accurately calculated
(a complete list of the phase space factors $G^{0\nu}(Q,Z)$ and $G^{2\nu}(Q,Z)$
can be found, e.g. in Ref. \cite{BV92}), 
$M^{0\nu}$ is the nuclear matrix element that can be
evaluated in principle, although with a considerable uncertainty
and is discussed in detail later, and finally the quantity
$\langle m_{\beta\beta} \rangle$ is the effective neutrino Majorana mass, representing
the important particle physics ingredient of the process.

In turn, the effective mass $\langle m_{\beta\beta} \rangle$ is related 
to the mixing angles $\theta_{ij}$ (or to the matrix elements $|U_{l,i}|$ of the
neutrino mixing matrix)
that are determined or constrained by the oscillation experiments, to the absolute neutrino
masses $m_i$ of the mass eigenstates $\nu_i$ and to the totally unknown additional
parameters, as fundamental as the mixing angles $\theta_{ij}$, 
the so-called Majorana phases $\alpha(i)$,
\begin{equation}
 \langle m_{\beta\beta} \rangle = | \Sigma_i |U_{ei}|^2 e^{i \alpha (i)} m_i | ~.
 \label{eq_mbb}
 \end{equation}
Here $U_{ei}$ are the matrix elements of the first row of the neutrino mixing matrix.

\begin{figure}
%\centerline{\psfig{file=rv-fig1.eps,width=3.8cm}}
\centerline{\psfig{file=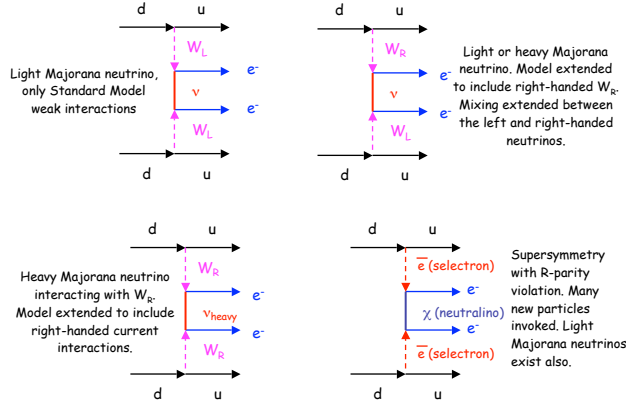,width=9.0cm}}
\caption{All these symbolic Feynman 
graphs potentially contribute to the  $0\nu\beta\beta$-decay
amplitude}
\label{fig:lnv}
\end{figure}

But that is not
the only possible mechanism. LNV interactions involving so far unobserved 
much heavier ($\sim$ TeV) particles
can lead to a comparable $0\nu\beta\beta$ decay rate. 
Some of the possible mechanisms of the elementary $dd \rightarrow uu+e^-e^-$
transition (the ``black box" in Fig. \ref{fig_SV})
are indicated in Fig. \ref{fig:lnv}.  Only the graph in the upper left
corner would lead to the usual relation between the decay rate and
neutrino mass, eq. \ref{eq_rate}. Thus, in the absence of additional
information about the mechanism responsible for the $0\nu\beta\beta$ decay
one could not unambiguously infer the magnitude of  $\langle m_{\beta\beta} \rangle$  
from the $0\nu\beta\beta$-decay rate.

In general $0\nu\beta\beta$ decay can be generated by (i) light massive Majorana 
neutrino exchange or (ii) heavy particle exchange (see, e.g. Refs.\cite{heavy,Pre03}),
resulting from LNV dynamics at some scale $\Lambda$ above the electroweak one.
The relative size of heavy ($A_H$) versus light 
particle ($A_L$) exchange contributions to the decay amplitude 
can be crudely estimated as follows~\cite{Mohapatra:1998ye}: 
\begin{equation}
A_L \sim G_F^2  \frac{\langle m_{\beta \beta} \rangle}{\langle k^2 \rangle}  ,~ 
 A_H \sim G_F^2  \frac{M_W^4}{\Lambda^5}  ,~
\frac{A_H}{A_L} \sim \frac{M_W^4 \langle k^2 \rangle } 
{\Lambda^5  \langle m_{\beta \beta} \rangle }  \ , 
\label{eq_estimate}
\end{equation}
where $\langle m_{\beta \beta} \rangle$ is the effective neutrino
Majorana mass, 
$\langle k^2 \rangle \sim ( 100 \ {\rm MeV} )^2 $ is the
typical light neutrino virtuality, and $\Lambda$ is the heavy
scale relevant to the LNV dynamics. 
Therefore,  $A_H/A_L \sim O(1)$ for  $\langle m_{\beta \beta} \rangle \sim 0.1-0.5$ 
eV and $\Lambda \sim 1$ TeV, and  thus the LNV dynamics at the TeV
scale leads to similar $0 \nu \beta \beta$-decay rate as the
exchange of light Majorana neutrinos with the effective mass 
$\langle m_{\beta \beta} \rangle \sim 0.1-0.5$ eV. 

Obviously, the lifetime measurement by itself
does not provide the means for determining the underlying mechanism.
The spin-flip and non-flip exchange can be, in principle,
distinguished by the measurement of the single-electron spectra or
polarization (see e.g. \cite{Doi}).  However, in most cases the
mechanism of light Majorana neutrino exchange, and of
heavy particle exchange, cannot be separated by the observation
of the emitted electrons. Thus one must look for other phenomenological
consequences of the different mechanisms. Here I discuss the
suggestion\cite{LNVus} that under natural assumptions the presence of low scale
LNV interactions, and therefore the absence of proportionality between
$\langle m_{\beta \beta} \rangle^2$ and the $0\nu\beta\beta$-decay
rate also affects muon lepton flavor violating (LFV)
processes, and in  particular enhances the $\mu \to e$ conversion 
compared to the $\mu \to e \gamma$ decay.

  The discussion is
concerned mainly with the branching ratios $B_{\mu \rightarrow e \gamma} = \Gamma
(\mu \rightarrow e \gamma)/ \Gamma_\mu^{(0)}$ and $B_{\mu \to e} =
\Gamma_{\rm conv}/\Gamma_{\rm capt} $, where $\mu \to e \gamma$ is
normalized to the standard muon decay rate $\Gamma_\mu^{(0)} = (G_F^2
m_\mu^5)/(192 \pi^3)$, while $\mu \to e$ conversion is normalized to
the muon nuclear capture rate $\Gamma_{\rm capt}$. The main diagnostic tool in our
analysis is the ratio 
\begin{equation}
{\cal R} = B_{\mu \to e}/B_{\mu \rightarrow e \gamma} ~,
\end{equation}
and the relevance of our observation relies on the potential
for LFV discovery in the forthcoming experiments  MEG~\cite{MEG}
($\mu \to e \gamma$) and MECO~\cite{MECO} 
($\mu \to e$ conversion)\footnote{Even though MECO 
experiment was recently cancelled, proposals
for experiments with similar sensitivity exist elsewhere.}.

The important quantities  are the scales for both 
LNV and LFV. If they are well above the weak scale, then one would not 
expect to observe any signal in the forthcoming LFV experiments, nor would 
the effects of heavy particle exchange enter $0\nu\beta\beta$ 
at an appreciable level. In this case, the only origin of a signal in 
$0\nu\beta\beta$ at the level of prospective experimental sensitivity 
would be the exchange of a light Majorana neutrino, leading to eq.(\ref{eq_rate}),
and allowing one to extract  $\langle m_{\beta \beta} \rangle$ from the decay rate.

In general, however, the two scales may be distinct, as in
SUSY-GUT~\cite{Barbieri:1995tw} or SUSY see-saw~\cite{Borzumati:1986qx} models. 
In these scenarios, both the Majorana neutrino mass as well as LFV effects are 
generated at the GUT scale.
The effects of heavy Majorana neutrino exchange in $0\nu\beta\beta$ are, thus, 
highly suppressed. In contrast,  the effects of GUT-scale LFV are transmitted 
to the TeV-scale by a soft SUSY-breaking sector without mass suppression 
via renormalization group running of the high-scale LFV couplings. 
Consequently, such scenarios could lead to observable effects 
in the upcoming LFV experiments, but with an ${\cal O}(\alpha)$ 
suppression of the branching ratio 
$B_{\mu\to e} $ relative to $B_{\mu\to e\gamma}$ 
due to the exchange of a virtual photon in the conversion process 
rather than the emission of a real one.

The case where the
scales of LNV and LFV are both relatively low ($\sim$ TeV)
is more subtle and requires more detailed analysis.
This is the scenario which might lead to observable signals
in LFV searches and at the same time generate ambiguities in
interpreting a  positive signal in $0 \nu \beta \beta$.
This is the case where one needs to develop some
discriminating criteria.

Based on the analysis in Ref. \cite{LNVus}, 
we can formulate the main conclusions regarding the discriminating
power of the ratio ${\cal R}$:
\begin{enumerate} 
\item 
Observation of both the LFV muon processes
$\mu \to e$ and $\mu \to e \gamma$ with relative ratio ${\cal R} \sim
10^{-2}$ implies, under generic conditions, that $\Gamma_{0 \nu \beta
\beta} \sim \langle m_{\beta \beta} \rangle^2$. Hence the relation
of the $0\nu\beta\beta$ lifetime to the absolute neutrino mass scale
is straightforward.
\item 
On the other hand, observation of LFV muon processes with
relative ratio ${\cal R} \gg 10^{-2}$ could signal non-trivial LNV
dynamics at the TeV scale, whose effect on $0 \nu \beta \beta$ has to
be analyzed on a case by case basis. Therefore, in this scenario no
definite conclusion can be drawn based on LFV rates.
\item
Non-observation of LFV in muon processes in forthcoming 
experiments would imply either that the scale of non-trivial LFV and
LNV is  above a few TeV, and thus 
$\Gamma_{0 \nu \beta\beta} \sim \langle m_{\beta \beta} \rangle^2$, 
or that any TeV-scale LNV is
approximately flavor diagonal (this is an important caveat).
\end{enumerate}

\section{Nuclear matrix elements}

Let us assume that the active neutrinos $\nu_1, \nu_2$ and $\nu_3$ are indeed 
massive Majorana fermions. If that is so then the neutrinoless $\beta\beta$ decay
will occur and its rate will be governed by eq.(\ref{eq_rate}). Thus, we need to know
the value of the nuclear matrix elements $M^{0\nu}$ in order to plan and interpret
the experiments. The nuclear transition involved consists of changing two
neutrons, bound in the ground state (always $J^{\pi} = 0^+$) of the initial
even-even nucleus, into two protons bound in the ground state 
(again always $J^{\pi} = 0^+$) of the final nucleus. (Here we do not consider the case
of the excited final nuclear states.)

\begin{figure}
%\centerline{\psfig{file=rv-fig1.eps,width=3.8cm}}
\centerline{\psfig{file=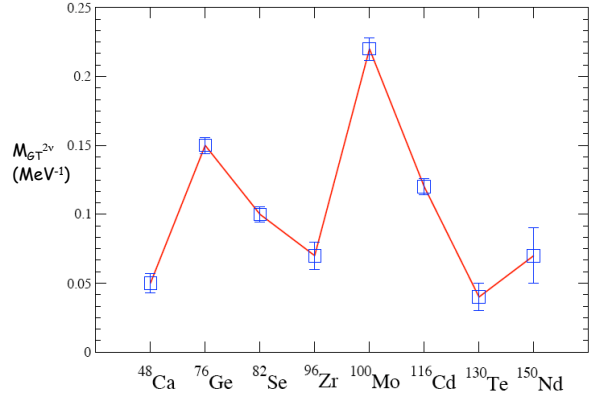,width=9.0cm}}
\caption{Nuclear matrix elements for the $2\nu\beta\beta$ decay extracted from the
measured half-lives.}
\label{fig_m2nu}
\end{figure}

In the $2\nu\beta\beta$ decay the two decaying neutrons are uncorrelated. The corresponding
momentum transfer from the initial neutron to the final proton is small ( $q \sim$ MeV, the
momentum of the $e^- + \bar{\nu}_e$, $qR \ll 1$) and thus the long wavelength approximation
is valid. Since the isospins of the initial and final nuclei are different, only the Gamow-Teller
operator remains and the corresponding nuclear matrix element is
\begin{equation}
M_{GT}^{2\nu} = \sum_m 
\frac{ \langle f || \sigma \tau_+ || m \rangle   \langle m || \sigma \tau_+ || i \rangle }
{E_m - (M_i + M_f)/2} ~,
\end{equation}
where $| m \rangle$ is the set of all $J^{\pi} = 1^+$ states in the virtual intermediate nucleus.
Once the lifetime of the $2\nu\beta\beta$ decay is known, the matrix element $M_{GT}^{2\nu}$
can be easily extracted since $1/T_{1/2} = G^{2\nu}(Q,Z) (M_{GT}^{2\nu})^2$, where the 
phase space factors $G^{2\nu}(Q,Z)$ can be easily and accurately calculated (see \cite{BV92}).
In Fig. \ref{fig_m2nu} the $M_{GT}^{2\nu}$ determined in this way are depicted. Note the rapid
variation of their value when different nuclei are involved.

For the $0\nu\beta\beta$ decay the situation is quite different. The two 
initial neutrons, that are transformed into the two final protons, are correlated. The virtual
massive Majorana neutrino that connects them has momentum at least of the order
of $q > 1/R$ ($R$ is the nuclear radius, as we will show below the actual momentum
transfer is even larger) and thus the long wavelength approximation
is not valid; all virtual intermediate states can in principle contribute. On the other hand,
the typical nuclear excitation are less than the energy of the virtual neutrino, hence
the ``closure approximation" is valid, and we usually do not
need to worry about the energies of intermediate states. 

In nuclear structure theory one begins with the mean field approximation in which the nucleons
are bound in a potential, but independent. That is, however, a poor approximation, and the
``residual interaction" need to be taken into account, typically with some truncation. 
There  are two basic and complementary methods of evaluating $M^{0\nu}$, the quasiparticle
random phase approximation (QRPA and its various generalizations) and the nuclear
shell model (NSM). These two methods differ fundamentally in the way the indicated
truncation is implemented. In QRPA one selects a wide interval of single-particle orbits 
but only a class of configurations (particle-hole and its iterations) of the nucleons
are taken into account. On the other hand, in NSM
only a relatively narrow  interval of single-particle orbits is chosen (one oscillator shell or less)
but all (or almost all) configurations of the valence nucleons residing on those orbits are included.
In both methods an effective interaction is used, based on the known nucleon-nucleon force,
but modified slightly using selected nuclear data as guidance. Since these two methods are
so different, it is important to test them against each other.

\begin{figure}
%\centerline{\psfig{file=rv-fig1.eps,width=3.8cm}}
\centerline{\psfig{file=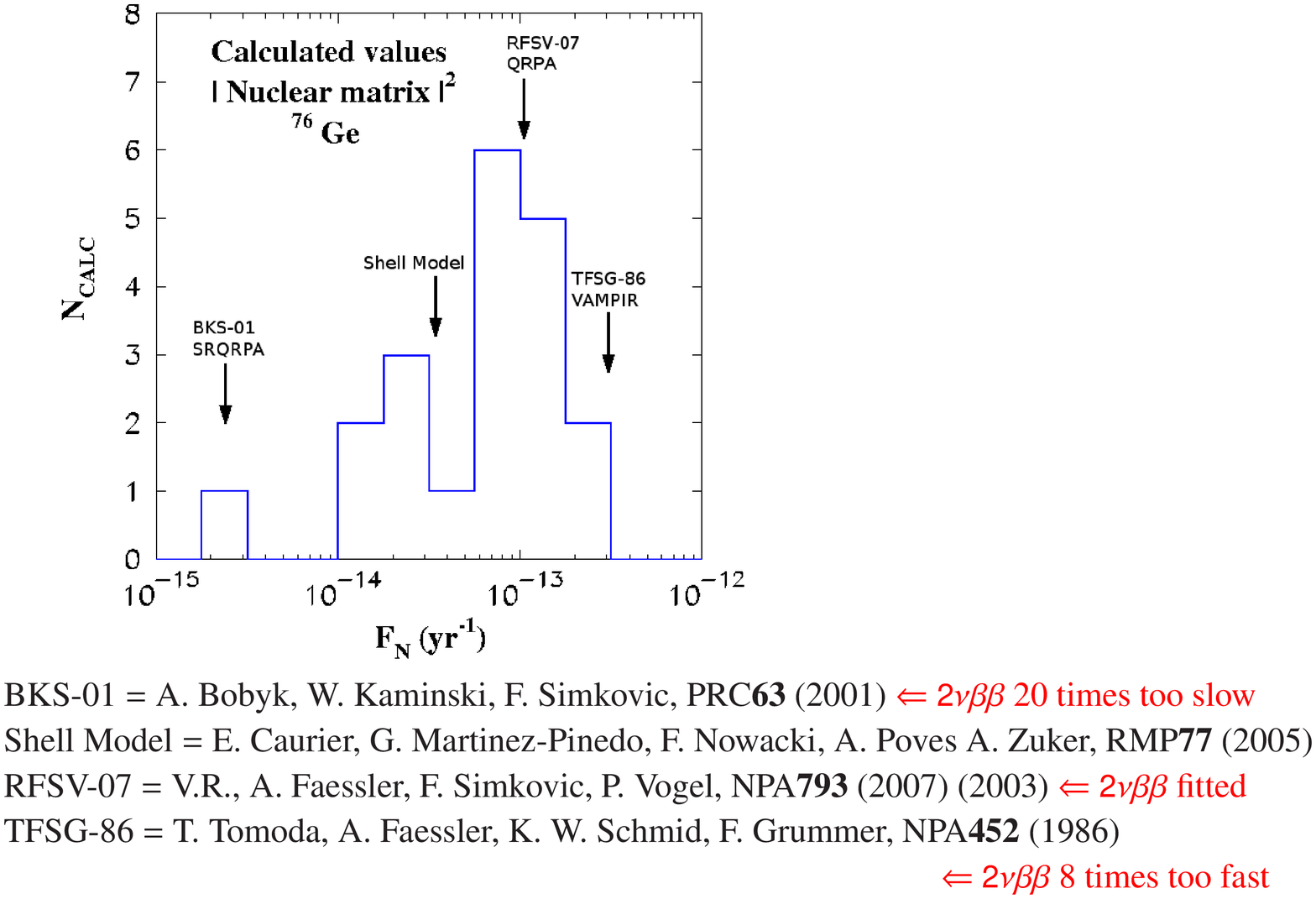,width=13.0cm}}
\caption{Histogram of older published calculated values of $(M^{0\nu})^2$ for $^{76}$Ge.
The failure of some of the calculations to reproduce the known $2\nu\beta\beta$-decay
lifetime is indicated. }
\label{fig_bah}
\end{figure}

There are many evaluations of the matrix elements $M^{0\nu}$ in the published literature
(for the latest review see \cite{AEE07}). However, the resulting matrix elements often do 
not agree with each other and it is difficult, based on the published material, to decide
who is right and who is wrong, and what is the theoretical uncertainty in $M^{0\nu}$.
That was stressed in the paper by Bahcall {\it et al.} few years ago \cite{Bahcall} where
a histogram of 20 calculated values of $(M^{0\nu})^2$ for $^{76}$Ge 
was plotted, with the implication
that the width of that histogram is a measure of uncertainty. That is clearly not a valid
conclusion as one could see in Fig. \ref{fig_bah} where the failure of the outliers
to reproduce the known $2\nu\beta\beta$-decay lifetime is indicated.

\begin{figure}[htb]
%\centerline{\psfig{file=rv-fig1.eps,width=3.8cm}}
\centerline{\psfig{file=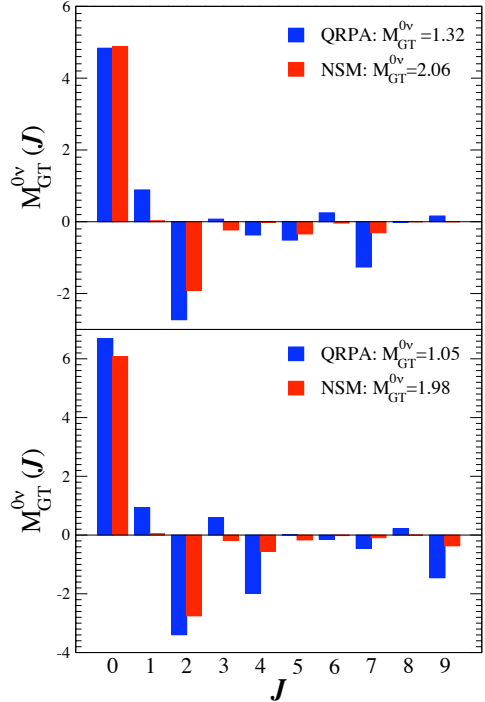,width=7.0cm}}
\caption{Contributions of different 
angular momenta
${\mathcal J}$ 
of the two participating neutrons 
to the Gamow-Teller part of $M^{0\nu}$ in $^{82}$Se (upper
panel) and $^{130}$Te (lower panel). The results of the nuclear
shell model (NSM) (dark
histogram) and QRPA treatments (lighter histogram) are
compared. Both calculations use the same single-particle spaces:
($f_{5/2},p_{3/2},p_{1/2}, g_{9/2}$) for $^{82}$Se and ($g_{7/2},
d_{5/2}, d_{3/2}, s_{1/2}, h_{11/2}$) for $^{130}$Te. 
In the QRPA calculation the experimental $2\nu\beta\beta$-decay 
rate was used to adjust the strength of the particle-particle interaction.}
\label{fig:competition}
\end{figure}

So why are the calculated values of different authors different?
In order to understand the difficulties in evaluating $M^{0\nu}$ we plot in Fig. \ref{fig:competition}
the contribution of different angular momenta ${\mathcal J}$ 
of the two transformed neutrons. There are two opposing tendencies in Fig.  \ref{fig:competition}.
The large positive contribution (essentialy the same in QRPA and NSM) is associated with the
so-called pairing interaction of neutrons with neutrons and protons with protons. As
the result of that interaction the nuclear ground state is mainly composed of Cooper-like
pairs of neutrons and protons coupled to ${\mathcal J} = 0$. The transformation of one 
neutron Cooper pair into one  Cooper proton pair is responsible for the ${\mathcal J} = 0$ piece in
 Fig.  \ref{fig:competition}. 

However, the nuclear hamiltonian contains, in addition, important neutron-proton interaction.
That interaction, primarily, causes presence in the nuclear ground state of ``broken pairs",
i.e. pairs of neutrons or protons coupled to ${\mathcal J} \ne 0$. Their effect, as seen in
Fig.  \ref{fig:competition}, is to reduce drastically the magnitude of $M^{0\nu}$. In treating
these terms, the agreement between QRPA and NSM is only semi-quantitative. Since the 
pieces related to the ``pairing" and ``broken pairs" contribution ale almost of the
same magnitude but of opposite signs, 
an error in one of these two competing tendencies is enhanced in the final  $M^{0\nu}$.
The competition, illustrated in Fig.  \ref{fig:competition}, is the main reason behind the
spread of the calculated $M^{0\nu}$. Many authors use different, and sometimes inconsistent,
treatment of the neutron-proton interaction.

In our QRPA (and RQRPA, renormalized QRPA) calculations \cite{us03,us06,us08}
we adjust the neutron-proton particle-particle interaction, responsible for the
``broken pairs" contribution, using the known $2\nu\beta\beta$-decay lifetimes, respectively
the corresponding $M^{2\nu}$ matrix elements. We based this procedure on the fact that
the Gamow-Teller strength, the contribution of the $1^+$ virtual intermediate states that is fully
responsible for the  $M^{2\nu}$, is the quantity most sensitive to the corresponding parameter,
usually denoted as $g_{pp}$. The nominal value, corresponding to the G-matrix based on
the realistic nucleon-nucleon force, is $g_{pp} = 1$. The renormalized $g_{pp}$ has values
between 0.8 and 1.2. 

In the NSM there is no analog to adjustment of the $g_{pp}$ parameter. The hamiltonian
(primarily its so-called monopole part) is adjusted so that a set of nuclear spectroscopic
data is optimally reproduced. That is a complicated and time consuming procedure, but
as a result a number of nuclear properties is well described. There is no attempt to reproduce
specifically the $2\nu\beta\beta$-decay lifetime, but the agreement with experiment is, in
most cases, acceptable \cite{sm95,sm07,sm08}.

To see some additional reasons why different authors obtain in their calculations different
nuclear matrix elements we need to analyze the dependence of the  $M^{0\nu}$ on the
distance $r$ between the pair of initial neutrons (and, naturally, the pair of final protons)
that are transformed in the decay process. That analysis reveals, at the same time, 
the various physics ingredients that must be included in the calculations so that 
realistic values of the $M^{0\nu}$ can be obtained.

The corresponding $0\nu\beta\beta$ decay operator contains, besides the spin and isospin
operators, the ``neutrino potentials", i.e. the Fourier transform of the neutrino propagator. The
simplest and most important of these potentials has the form
\begin{equation}
H(r) = \frac{R}{r} \Phi(\omega r) ~,
\end{equation}
where $R$ is the nuclear radius introduced here to make the potential, and the resulting 
$M^{0\nu}$, dimensionless (the  $1/R^2$ in the phase space factor compensates
for this), $r$ is the distance between the transformed neutrons (or protons) and
$\Phi(\omega r) $ is a rather slowly varying function of its argument.

\begin{figure}
%\centerline{\psfig{file=rv-fig1.eps,width=3.8cm}}
\centerline{\psfig{file=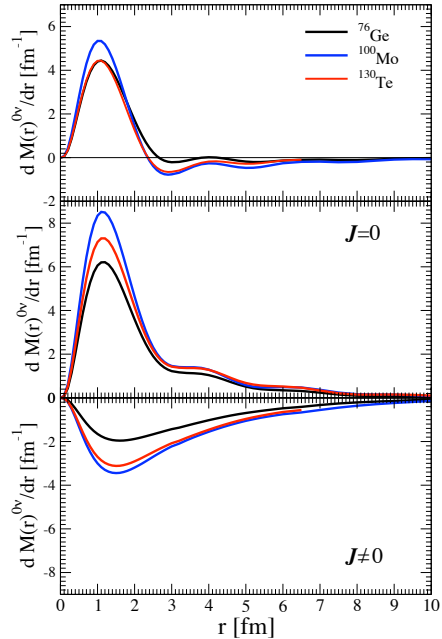,width=6.5cm}}
\caption{The dependence on $r$ of $M^{0\nu}$ for
$^{76}$Ge, $^{100}Mo$ and $^{130}$Te. The upper panel shows the full
matrix element, and the lower panel shows separately `pairing'
(${\mathcal J} = 0$ of the two participating neutrons) 
and `broken pair' (${\mathcal J} \ne 0$)
contributions.}
\label{fig:radial}
\end{figure}

From the form of the potential $H(r)$ one would, naively, expect that the 
characteristic value of $r$ is the typical distance between the nucleons in a nucleus, namely
that $\bar{r} \sim R$. However, that is not true as was demonstrated first in Ref. \cite{us08}
and illustrated in Fig. \ref{fig:radial}. One can see there that the competition between the
``pairing" and ``broken pairs" pieces essentially removes all effects of $r \ge 2-3$ fm.
Only the relatively short distances contribute significantly. The same result was
obtained in the NSM \cite{sm08}. (We have also shown in \cite{us08} that 
an analogous result is obtained in an exactly solvable, semirealistic model. There
we also showed that this behaviour is restricted to an interval of the parameter $g_{pp}$
that contains the realistic value near unity.)   

Once the $r$ dependence displayed in Fig. \ref{fig:radial} is accepted, several new
physics effects come to mind. One of them is the short-range nucleon-nucleon repulsion
known from scattering experiments. Two nucleon strongly repel each other at distances
$r \le 0.5-1.0$ fm, i.e. the distances very relevant to evaluation of the $M^{0\nu}$. The
nuclear wave functions used in QRPA and NSM, products of the mean field single-nucleon
wave function, do not take into account the influence of this repulsion that is irrelevant
in most standard nuclear structure theory applications. The standard way to include the
effect is to modify the radial dependence of the $0\nu\beta\beta$ operator so that the
effect of short distances (small values of $r$) is reduced.

\begin{figure}
%\centerline{\psfig{file=rv-fig1.eps,width=3.8cm}}
\centerline{\psfig{file=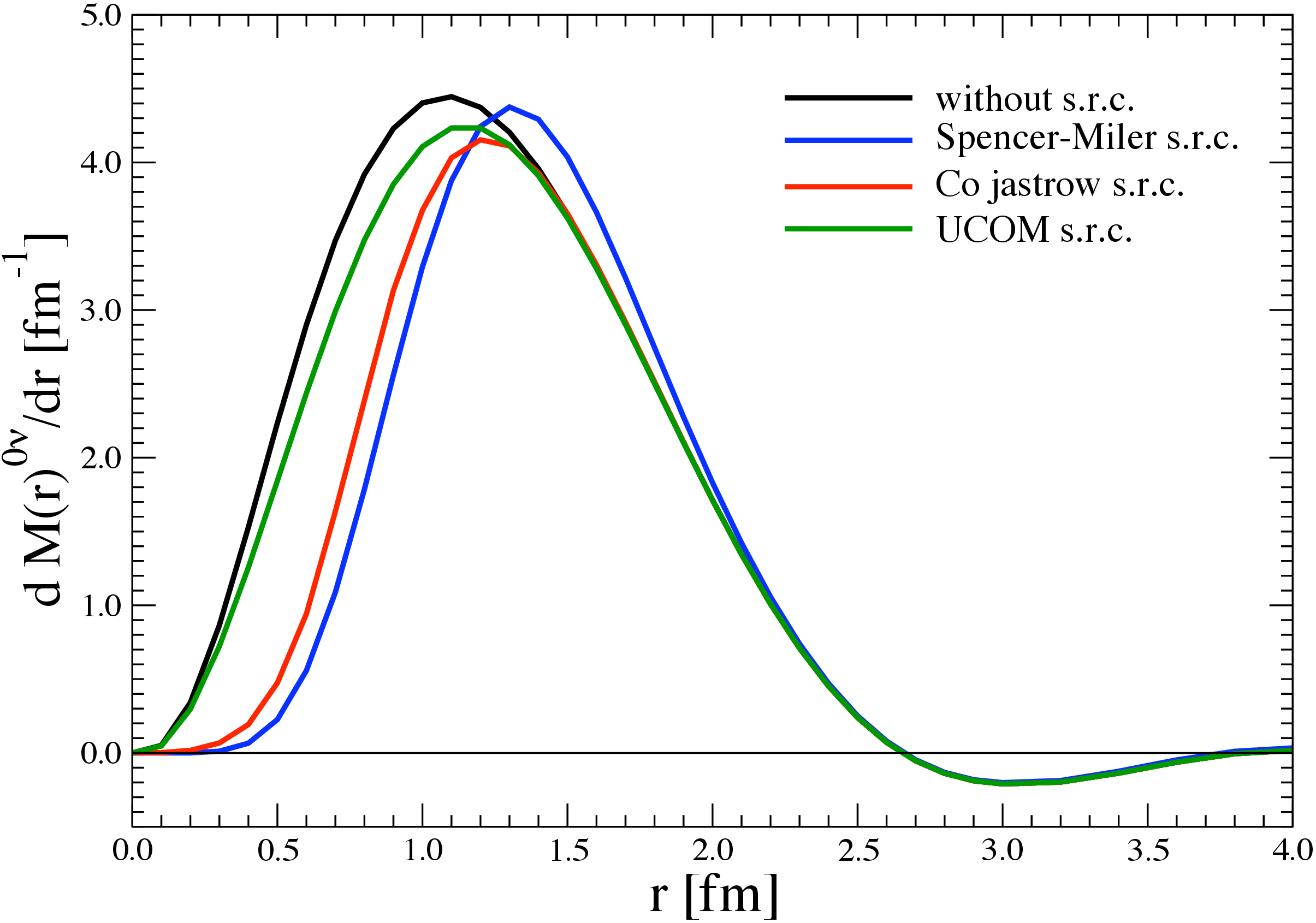,width=9.0cm}}
\caption{The $r$ dependence of $M^{0\nu}$ in
$^{76}$Ge. The four curves show the effects of different treatments
of short-range correlations. The resulting $M^{0\nu}$ values are
5.32 when the effect is ignored, 5.01 when the UCOM transformation
%\cite{UCOM} 
is applied and 4.14 when the treatment based on the
Fermi hypernetted chain 
%\cite{Co}
and 3.98 when the phenomenological Jastrow 
function is used. (See the text for details.)
% \cite{Spencer}.
}
\label{fig:src}
\end{figure}

Traditionally, this was done by multiplying the neutrino potential by  
the square of a  Jastrow-like function first derived in \cite{Spencer} and in
a more modern form in \cite{Co}. That phenomenological procedure reduces 
the magnitude of   $M^{0\nu}$ by 20-25\% as illustrated in Fig. \ref{fig:src}.
Recently, another procedure, based on the Unitary Correlation Operator Method
(UCOM) has been proposed \cite{UCOM}. That procedure, still applied not fully consistently,
reduces the $M^{0\nu}$  much less, only by about 5\% \cite{Kort}. It is prudent to include
these two possibilities as extremes and the corresponding range as systematic error.
Once a consistent procedure is developed, consisting of deriving an effective $0\nu\beta\beta$
decay operator that includes (probably perturbatively) the effect of the high momentum
(or short range) that component of the systematic error could be substantially reduced.  

Another effect that needs to be taken into account is the nucleon finite size. That is included,
usually, by introducing the dipole form of the nucleon form factor
\begin{equation}
f_{V,A} =\left(  \frac{1}{1 + q^2/M_{V,A}^2} \right)^2 ~,
\end{equation}
where the cut-off parameters $M_{V,A}$ have values (deduced in the reactions of
free neutrinos with free or quasifree nucleons) $\sim$ 1 GeV. This corresponds to
the nucleon size of $\sim 0.5-1.0$ fm. Note that in our case we are dealing
with neutrinos far off mass shell, and bound nucleons, hence it is not obvious
that the above form factors are applicable. It turns out, however, that
once the short range correlations are properly included (by either of the procedures 
discussed above) the $M^{0\nu}$ becomes essentially independent of the adopted
values when $M_{V,A} \ge$ 1 GeV. However, in the past various authors neglected the
effect of short range correlations, and in that case a proper inclusion of nucleon form factor
(or their neglect) again causes variations in the calculated $M^{0\nu}$ values.

Yet another correction that various authors neglected must be included in a correct treatment.
Since $r \le$ 2-3 fm is the relevant distances, the corresponding momentum transfer $1/r$
is of the order of $\sim$200 MeV, much larger than in the ordinary $\beta$ decay. Hence the induced
nucleon currents, in particular the pseudoscalar (since the neutrinos are far off mass shell)
give noticeable contributions \cite{us08,sim99}.

We have, therefore, identified the various physics effects that ought to be included in a
realistic evaluation of $M^{0\nu}$ values. The spread of the calculated values, noted
by Bahcall {\it et al.} \cite{Bahcall} can be often attributed to the fact that various authors either
neglect some of them, or include them inconsistently.

\section{Discussion and conclusion}

\begin{figure}
%\centerline{\psfig{file=rv-fig1.eps,width=3.8cm}}
\centerline{\psfig{file=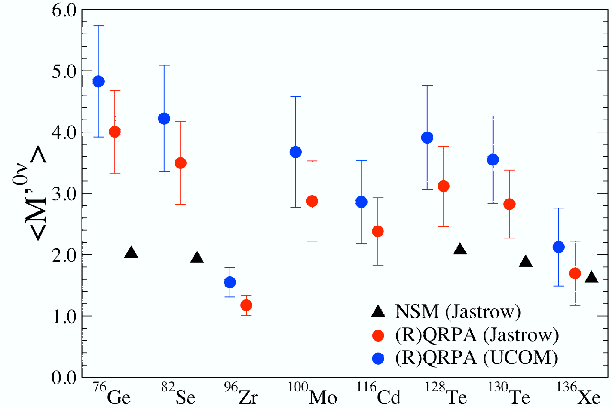,width=9.0cm}}
\caption{The full ranges of $M^{'0\nu}$ with the two alternative
treatments of the short range correlations. For comparison
the results of a recent Large Scale Shell Model evaluation of
$M^{'0\nu}$ that used the Jastrow-type treatment of short range
correlations are also shown (triangles).}
\label{fig:me}
\end{figure}

Even though we were able to explain, or eliminate, a substantial part of the spread
of the calculated values of the nuclear matrix elements, sizeable systematic uncertainty
remains. That uncertainty, within QRPA and RQRPA, as discussed in Refs.\cite{us03,us06},
is primarily related to the difference between these two procedures, to the
size of the single-particle space included, whether the so-called quenching of the axial current
coupling constant $g_A$ is included or not, and to the systematic error in the treatment
of short range correlations \cite{us08}. In Fig. \ref{fig:me} the full ranges of the resulting
matrix elements $M^{0\nu}$ is indicated. The indicated error bars are highly correlated; e.g.,
if true values are near the lower end in one nucleus, they would be near the lower ends in all
indicated nuclei.

The figure also shows the most recent NSM results \cite{sm07}. Those results, obtained with
Jastrow type short range correlation corrections, are noticeably lower than the QRPA values.
That difference is particularly acute in the lighter nuclei $^{76}$Ge and $^{82}$Se. While
the QRPA and NSM agree on many aspects of the problem, in particular on the role
of the competition between ``pairing" and ``broken pairs" contributions and on the
$r$ dependence of the matrix elements, the disagreement in the actual values remains
to be explained. 

When one compares the $2\nu$ and $0\nu$ matrix elements (Figs. \ref{fig_m2nu} and \ref{fig:me})
the feature to notice is the fast variation in $M^{2\nu}$ when going from one nucleus
to another while $M^{0\nu}$ change only rather smoothly, in both QRPA and NSM. This 
is presumably related to the high momentum transfer (or short range) involved in
$0\nu\beta\beta$. That property of the $M^{0\nu}$ matrix elements makes the comparison
of results obtained in different nuclei easier and more reliable.

While substantial progress has been achieved, we are still somewhat far from being able to evaluate
the $0\nu\beta\beta$ nuclear matrix elements  confidently and accurately. Perhaps at the
next Venice meeting we will be able to report that we reached that goal.

\section{Acknowledgments}
 The original results reported here were obtained in collaboration with Fedor \v{S}imkovic,
 Vadim Rodin, Amand Faessler and Jonathan Engel. The fruitful collaboration with them
 is gratefully acknowledged. Part of the work was performed at MPI-K, Heidelberg; the
 hospitality of Prof. M. Lindner is deeply appreciated.

\end{document}